\documentclass[prl,aps,amssymb,showpacs,twocolumn]{revtex4}
\usepackage{amsmath}
\usepackage{amssymb}
\usepackage{amsthm}
\usepackage{amsfonts}
\usepackage{algorithmic}
\usepackage{enumerate}
\usepackage{latexsym}
\usepackage[dvips]{graphicx}

\newcommand{\beq}{\begin{equation}}
\newcommand{\eneq}{\end{equation}}

\input{epsf}

\begin{document}

\tolerance 10000


\title{Magnetic Instability in Strongly Correlated Superconductors}

\author { Bogdan A. Bernevig$^\dagger$, Robert B. Laughlin$^\dagger$ and
David I. Santiago$^{\dagger, \star}$ }

\affiliation{ $\dagger$ Department of Physics, Stanford
University,
         Stanford, California 94305 \\ $\star$ Gravity Probe B Relativity
Mission, Stanford, California 94305}
\begin{abstract}
\begin{center}

\parbox{14cm}{Recently a new phenomenological Hamiltonian has been proposed to 
describe the superconducting cuprates. This so-called Gossamer Hamiltonian is 
an apt model for a superconductor with strong on-site Coulomb repulsion between
the electrons. It is shown that as one approaches half-filling the Gossamer 
superconductor, and hence the superconducting state, with strong repulsion is 
unstable toward an antiferromagnetic insulator an can undergo a quantum phase 
transition to such an insulator if one increases the on-site 
Coulomb repulsion.}

\end{center}
\end{abstract}

\pacs{74.20.-z, 74.20.Mn, 74.72.-h, 71.10.Fd, 71.10.Pm }

\maketitle

The parent materials of the high temperature superconducting cuprates are 
correlated antiferromagnetic insulators. When they are
half-filled, with  one hole per Copper, they insulate despite having an odd 
number of electrons in their ``valence band''. The antiferromagnetism and 
insulation stem from the strong on-site Coulomb repulsion amongst the Copper 
d-electrons. These electron correlations have been postulated to be essential 
to the superconductivity in the cuprates\cite{phil} since its discovery. 

We want to suggest that the correlation effects might be seducing us into 
misidentifying them as the key ingredient  for high T$_c$ superconductivity. 
In order to study the consequences and viability of such an idea we study a 
Hamiltonian recently proposed by
one of us\cite{bob} which has a d-wave superconducting ground state 
{\it for all dopings} up to the half-filled undoped state. This superconductor 
was baptized the Gossamer superconductor.

The correlation and magnetic effects compete and are detrimental to the 
superconductivity. In previous work\cite{bob} it was estimated that, at
strong projection, the spectral function will evolve with decreasing doping 
toward that of an insulator with two  Hubbard bands, a Hubbard gap and an ever 
fainter redistribution of spectral weight to mid-gap states\cite{fuji} 
corresponding to the collapsing superfluid density. 

Thus superconductors with strong on-site repulsion are 
{\it spectroscopically identical} to so-called doped Mott-insulators close to 
half-filling, except for a small amount of conducting fluid corresponding to 
the dephased superconductor. This naturally accommodates
experiments that hint at conduction in the supposedly antiferromagnetic
insulating phase\cite{mottcond} and the existence of a d-wave node deep in the 
underdoped regime\cite{fuji}.

We identify the pseudogap\cite{timusk} measured in underdoped cuprates with
the Cooper pairing gap. In this region the superconducting transition 
temperature, T$_c$, is lower than the pairing temperature because the
Gossamer superconductor is becoming ever increasingly unstable to loss of
phase coherence due to the small superfluid density\cite{emery,dynes}.

\begin{figure}
\input{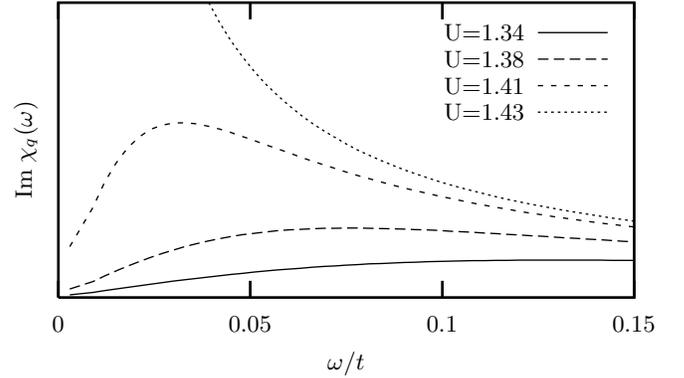}
\caption{Dependence of the imaginary part of the spin susceptibility in
RPA approximation on the energy in units of $t$. The specific curves
plotted here are for $\Delta_0 = 0.4 t$ and $q=(\pi,\pi)$. Upon
increasing the Hubbard U toward the critical value U$=1.43t$ we notice the
divergence of the susceptibility, a sign that magnetic order is about to
set in.} \label{fig}
\end{figure}

In the present note we show that superconducting state with strong on-site 
repulsion is unstable toward insulation and antiferromagnetism close to
half-filling by studying such a half-filling instability in the the Gossamer 
superconductor. For the Gossamer superconductor the instability is exactly at
half-filling while for a different Hamiltonian the instability can occur
at nonzero doping. For example, antiferromagnetic or stripe 
ground states\cite{tranquada} can be
stabilized by adding an extra Hubbard $U$ term to the Gossamer Hamiltonian.

The Gossamer superconductor is defined as a superconducting
ground state which contains Coulomb correlations. These are introduced by
a partial
Gutzwiller projection which decreases the probability of having
two electrons on the same site:

\begin{equation}
\Pi_{\alpha_0} = \prod_j z^{(n_{j \uparrow} + n_{j \downarrow}) /
2}_{0}(1 - \alpha_0 n_{j \uparrow} n_{j \downarrow}) \; \; \; .
\end{equation}
\noindent $ 0 \le \alpha_0 < 1 $ is a measure of how effective the
projector is and in a real material it will be related to the
Coulomb repulsion. The factor of $z_0$, the quantum fugacity, in
the projector is the extra probability of having an electron at
site $j$ after projecting and is necessary in order to keep the
total number of particles constant at $(1-\delta)N$ after
projecting. The fugacity is given by $z_0 = (\sqrt{1 - \alpha(1-
\delta^2)} - \delta)/ [(1-\alpha)(1-\delta)]$\cite{bob} with
$(1-\alpha_0)^2 = 1 -\alpha$.

The Gossamer superconducting ground state is postulated to be

\begin{equation}
| \Psi  > = \Pi_\alpha \; | \Phi  > \label{gs} \; \; \; .
\end{equation}

\noindent Here $| \Phi >$ is the BCS ground state:

\begin{equation}
|\Phi > = \prod_{\vec{k}} (u_{\vec{k}} + v_{\vec{k}}
c^{\dagger}_{\vec{k} \uparrow} c^{\dagger}_{-\vec{k} \downarrow})
|0> \; \; \; .
\end{equation}

\noindent where $u_{\vec{k}}$, $v_{\vec{k}}$ are the well-known BCS
pairing amplitudes given by:

\begin{equation}
u_{\vec{k}}=\sqrt{\frac{E_{\vec{k}} + \epsilon_{\vec{k}} - \mu }{2
E_{\vec{k}}}} \; \; v_{\vec{k}}=\sqrt{\frac{E_{\vec{k}} -
(\epsilon_{\vec{k}}
- \mu) }{2E_{\vec{k}}}} \; \; \; .
\end{equation}
\noindent with dispersion $E_{\vec{k}}= \pm \sqrt{(\epsilon_{\vec{k}} - 
\mu)^2
+ \Delta^{2}_{\vec{k}}}$  where $\epsilon_{\vec{k}}$ is the kinetic energy, 
$\mu$ is the chemical potential and $\Delta_{\vec{k}}$ is the
superconducting gap. We take $\epsilon_{\vec{k}}= 2t (\cos(k_x a) + 
\cos(k_y a))$ for a square lattice with spacing $a$, and 
$\Delta_{\vec{k}}= \Delta_o ( \cos (k_x a) - \cos(k_ya))$
for a d-wave gap as found for the superconducting cuprates\cite{timusk}. In 
superconductors the coherence factors $u_{\vec{k}}$ and
$v_{\vec{k}}$ are related to the number of carriers in order to
set the value of the chemical potential. For doped cuprates we
have $\frac{1}{N} \sum_{\vec{k}} v_{\vec{k}}^2 = 1 - \frac{1}{N}
\sum_{\vec{k}} u_{\vec{k}}^2 = (1 - \delta) / 2$ where $\delta$ is the
doping level.

Projected ground states like the Gossamer ground state have been
previously used in the literature\cite{rand} to describe high
temperature superconductors. We only consider projection away from
full projection ($\alpha < 1$) in order for the partial projector
to have an inverse:
\begin{equation}
\Pi^{-1}_{\alpha} = \prod_j z^{-(n_{j \uparrow} + n_{j
\downarrow}) / 2}_{0}(1 + \beta_0 n_{j \uparrow} n_{j \downarrow})
\; \; \; ,
\end{equation}
\noindent with $\beta_0 = \alpha_0 / (1 - \alpha_0)$. By virtue of
this invertibility, the Gossamer ground state is adiabatically continuable 
to
the BCS ground state and its uniqueness follows from the uniqueness of the BCS
ground state up to a phase. Therefore the Gossamer superconductor describes 
the
same phase of matter as the BCS superconductor.

The Gossamer ground state is the exact ground state of the Gossamer
Hamiltonian:

\begin{equation}
{\cal H} = \sum_{\vec{k} \sigma} E_{\vec{k}} B_{\vec{k}
\sigma}^\dagger B_{\vec{k} \sigma}, \;\; \; \; B_{\vec{k} \sigma}
|\Psi \rangle =0 . \label{gossham}
\end{equation}
\noindent where:

\begin{displaymath}
B_{\vec{k} \uparrow \{\downarrow\}} = \Pi_\alpha b_{\vec{k}
\uparrow \{\downarrow\}} \Pi_\alpha^{-1} = \frac{1}{\sqrt{N}}
\sum_j^N e^{i \vec{k} \cdot \vec{r}_j}
\end{displaymath}

\begin{equation}
\times \biggl[ z_0^{-1/2} u_{\vec{k}} (1 + \beta_0 n_{j \downarrow
\{\uparrow\}} ) c_{j \uparrow \{\downarrow\}} \pm z_0^{1/2}
v_{\vec{k}} (1 - \alpha_0 n_{j \uparrow \{\downarrow \}} ) c_{j
\downarrow \{\uparrow\}}^\dagger \biggr] 
\end{equation}
\noindent with $b_{\vec{k}\uparrow \{\downarrow\}} =  u_{\vec{k}}
c_{j \uparrow \{\downarrow\}} \pm v_{\vec{k}} c_{j \downarrow
\{\uparrow\}}^\dagger$ the Bogoliubov quasiparticle operators. Note that
the Gossamer Hamiltonian is a supersymmetric Hamiltonian in the
sense that it annihilates the ground state and it is a nonnegative operator.

A superconductor with strong on-site Coulomb repulsion is described by the
Gossamer Hamiltonian with nearly full projection, i.e. $\alpha_0 \rightarrow
1^-$. The strong projector collapses the superfluid density with doping
according to $2\delta /(1 + \delta)$\cite{bob} and introduces new 
correlations and instabilities generic to other types of ordering as we show 
in the present letter.

In order to determine the correlations arising in the
different limits of the Gossamer Hamiltonian Eq. (\ref{gossham}),
we will expand the Hamiltonian and analyze its terms.
After some manipulation, we can bring the Hamiltonian in the form
of a sum of 3 physically distinct terms:
\beq {\cal{H}} =
\sum_{\vec{k} \sigma} E_{\vec{k}} B^\dagger_{\vec{k} \sigma}
B_{\vec{k} \sigma} = \cal{A} + \cal{B} + \cal{C}
\eneq \noindent
where $\cal{A}, \cal{B}, \cal{C}$ are, explicitly:
\begin{displaymath} {\cal{A}}= \sum_{\vec{k}} \frac{E_{\vec{k}}}{N} 
\sum_{i,j}^N e^{-i \vec{k}
(\vec{r}_i -\vec{r}_j)} \{z_0^{-1} u_{\vec{k}}^2 (1+\beta_0
n_{i\downarrow})(1+\beta_0 n_{j\downarrow}) c^\dagger_{i \uparrow}
c_{j \uparrow} +
\end{displaymath}
\begin{equation}
z_0 v_{\vec{k}}^2 (1-\alpha_0 n_{i \downarrow})(1-\alpha_0 n_{j
\downarrow}) c_{i \uparrow} c^\dagger_{j \uparrow}\} +
\{\uparrow\rightleftarrows\downarrow \}
\end{equation}

\begin{displaymath}
{\cal{B}}=\sum_{\vec{k}} \frac{E_{\vec{k}}}{N} \sum_{i,j}^N
e^{-i\vec{k}(\vec{r}_i - \vec{r}_j)} u_{\vec{k}} v_{\vec{k}} \{
(1+\beta_0 n_{i \downarrow})(1-\alpha_0 n_{j \uparrow})
c^\dagger_{i \uparrow} c^\dagger_{j\downarrow} -
\end{displaymath}
\begin{equation}
+(1+\beta_0 n_{j \downarrow})(1-\alpha_0 n_{i \uparrow}) c_{i
\downarrow} c_{j\uparrow} \}-\{\uparrow\rightleftarrows\downarrow
\} \label{B}
\end{equation}

\begin{displaymath}
{\cal{C}}=\sum_{\vec{k}} \frac{E_{\vec{k}}}{N}  \sum_{j}^N
u_{\vec{k}} v_{\vec{k}} \{ \alpha_0 (1+\beta_0 n_{j \downarrow})
c^\dagger_{j \uparrow} c^\dagger_{j \downarrow}
\end{displaymath}

\begin{equation}
+ \beta_0 (1-\alpha_0 n_{j \uparrow}) c_{j \downarrow} c_{j
\uparrow} \} - \{\uparrow\rightleftarrows\downarrow \}
\end{equation}

\noindent The last term $\cal{C}$ vanishes by use of the identity:

\beq 
E_{\vec{k}} u_{\vec{k}} v_{\vec{k}} = \frac{\Delta_{\vec{k}}}{2} 
\eneq 

\noindent as well as the relation:

\beq \sum_{\vec{k}} e^{-i\vec{k}(\vec{r}_i - \vec{r}_j)}
\Delta_{\vec{k}} = \left\{
\begin{array}{ccc}
  0 & , & \vec{r}_i \ne \vec{r}_j  +\vec{a}\\
  \Delta_0 & , & \vec{a} \parallel \hat{x} \\
  -\Delta_0 & , & \vec{a} \parallel \hat{y} \\
\end{array} \right\}\eneq \noindent
(true for a d-wave gap) where $\vec{a}$ is vector pointing toward
a nearest neighbor in the lattice. The $\cal{A}$ term is
responsible for the chemical potential, kinetic energy, as well as a
Hubbard U terms. $\cal{B}$ is responsible for the
super-conducting part of the Gossamer Hamiltonian. The
d-wave form of the gap makes $\cal{B}$ have no on-site
contributions. $\cal{A}$ contains on-site and off-site
contributions.

We can write $\cal{A}$ as a sum between on-site and off-site
contributions $ {\cal{A}} = {\cal{A}}_{on
\; site} + {\cal{A}}_{off\;site}$, where the two contributions read:

\begin{displaymath} {\cal{A}}_{on
\; site}= \sum_{\vec{k}} \frac{E_{\vec{k}}}{N} \sum_{j}^N  \{z_0^{-1}
u_{\vec{k}}^2
(1+\beta_0 n_{j\downarrow})^2 c^\dagger_{j \uparrow} c_{j \uparrow} +
\end{displaymath}

\begin{equation} + z_0 v_{\vec{k}}^2 (1-\alpha_0 n_{j \downarrow})^2
c_{j \uparrow} c^\dagger_{j \uparrow} \} +\{ \uparrow
\rightleftarrows \downarrow \}
\end{equation} \noindent

\begin{displaymath} {\cal{A}}_{off\;site}
=\sum_{\vec{k}} \frac{E_{\vec{k}}}{N} \sum_{i \ne j}^N e^{-i
\vec{k} (\vec{r}_i -\vec{r}_j)}
\end{displaymath}

\begin{displaymath}
\times \{z_0^{-1} u_{\vec{k}}^2 (1+\beta_0
n_{i\downarrow})(1+\beta_0 n_{j\downarrow}) c^\dagger_{i \uparrow}
c_{j \uparrow} +
\end{displaymath}

\begin{equation}
+z_0 v_{\vec{k}}^2 (1-\alpha_0 n_{i
\downarrow})(1-\alpha_0 n_{j \downarrow})  c^\dagger_{j \uparrow}
c_{i \uparrow} \} + \{\uparrow \rightleftarrows \downarrow \}
\end{equation}

\noindent The Hubbard U term will arise out of the on-site contribution,
${\cal{A}}_{on \; site}$. After some operator algebra the term can
be transformed into:

\begin{displaymath} {\cal{A}}_{on
\; site}
= \sum_{\vec{k}} \frac{E_{\vec{k}}}{N} \sum_j^N \{ 2 z_0
v^2_{\vec{k}} +
\end{displaymath}

\begin{displaymath}
+[z_0^{-1} u^2_{\vec{k}} - z_0 v^2_{\vec{k}} -2\alpha_0 z_0
v^2_{\vec{k}} + \alpha_0^2 z_0 v^2_{\vec{k}}] (n_{j \uparrow} +
n_{j \downarrow}) +
\end{displaymath}

\beq +[z_0^{-1} u^2_{\vec{k}}(4\beta_0 + 2 \beta_0^2) + z_0
v^2_{\vec{k}}(4\alpha_0 - 2\alpha_0^2)]n_{j \uparrow}  n_{j
\downarrow}\}
\eneq

\noindent The first term is a zero-point energy, the second term
is a chemical potential, and, most interestingly, the third term
is the Hubbard U term ($\sum_j^N U n_{j \uparrow}  n_{j \downarrow} $)
with:

\begin{equation}
U=\sum_{\vec{k}} \frac{E_{\vec{k}}}{N} [z_0^{-1}
u^2_{\vec{k}}(4\beta_0 + 2 \beta_0^2) + z_0
v^2_{\vec{k}}(4\alpha_0 - 2\alpha_0^2)]
\end{equation}

\noindent Thus the Gossamer Hamiltonian has a Hubbard $U$ term.

The Gossamer Hamiltonian is constructed such that its ground state is 
superconducting for all nonzero dopings. Hence it will be most susceptible to
other kinds of order, antiferromagnetism in the case at hand, at zero doping 
where the superfluid density has collapsed to zero. We thus concentrate on 
half-filling $\delta =0$, where $U$ becomes: 

\beq 
U|_{\delta = 0}= \sum_{\vec{k}} \frac{2
E_{\vec{k}}}{N} \frac{\alpha_0(2- \alpha_0)}{1-\alpha_0} \label{U}
\eneq 

\noindent As we can see, at almost full projection $\alpha_0
\rightarrow 1^-$, $U$ becomes very large.

The off-site contributions of $\cal{A}$ give hopping (kinetic)
term in the Hamiltonian. Prior to the partial Gutzwiller
projection ($\alpha_0 =0$, $z_0 =1$) this term is just the kinetic energy or
hopping term of the Hamiltonian
$\sum_{\vec{k} \sigma} (\epsilon_{\vec{k}} - \mu)
c^{\dagger}_{\vec{k} \sigma} c_{\vec{k} \sigma} $. Including the
partial projection, particularizing to zero doping and
{\it imposing} the mean field values $\langle n_{i\uparrow} \rangle = 
\langle n_{i\downarrow} \rangle = 1/2 $, the off-site, after some manipulation,
becomes:

\beq {\cal{A}}_{off\;site} =\frac{1}{4}
\frac{(2-\alpha_0)^2}{(1-\alpha_0)} \sum_{\vec{k} \sigma}
(\epsilon_{\vec{k}} - \mu) c^\dagger_{\vec{k} \sigma} c_{\vec{k}
\sigma}  \label{t}
\eneq

\noindent At half filling, the effect of
the partial projection on the kinetic term in the Gossamer
Hamiltonian is, surprisingly, just  a renormalization. Upon strong projection, 
the physically 
relevant ratio, $U/t$, approaches a number of order unity or greater which 
provides the right physics for the appearance of strong antiferromagnetic 
correlations and the
opening of an insulating gap\cite{fz}. 

The superconducting part of the Gossamer Hamiltonian, $\cal{B}$, given
in Eq. (\ref{B}) is, when unprojected, just the pair attraction term
$\sum_{\vec{k}} \Delta_{\vec{k}} [
c^{\dagger}_{\vec{k} \uparrow} c^{\dagger}_{-\vec{k} \downarrow} +
c_{-\vec{k} \downarrow} c_{\vec{k} \uparrow} ]$ from the mean
field d-wave superconducting Hamiltonian. Concentrating on half-filling, we
estimate ${\cal{B}}$ by {\it imposing} the mean field condition
$\langle n_{i\uparrow} \rangle = \langle n_{i \downarrow} \rangle =1/2$, and
keeping in mind that $E_{\vec{k}}$ and
$\Delta_{\vec{k}}$ are even in $\vec{k}$. We thus obtain

\beq {\cal{B}}=\frac{1}{4}
\frac{(2-\alpha_0)^2}{1-\alpha_0} \sum_{\vec{k}} \Delta_{\vec{k}}
 (c^\dagger_{\vec{k} \uparrow}
c^\dagger_{-\vec{k} \downarrow} + c_{-\vec{k} \downarrow}
c_{\vec{k} \uparrow})
\eneq

\noindent The new superconducting gap, at half-filling, upon projection is 
still d-wave, 
and is renormalized by the same constant as the kinetic energy. Upon strong
projection, the physically 
relevant ratio, $U/ \Delta_0$, is a number of order unity or greater, the
right physics for antiferromagnetism and insulation. It is  very 
interesting that the gap survives 
along with Hubbard $U$ term at half-filling where the superfluid density 
is zero.

We have thus shown that at half-filling and under strong projection the
Gossamer superconductor Hamiltonian is a Hubbard Hamiltonian with a d-wave
pairing interaction added to it. If we define the ``spinors''

\begin{equation}
\Psi_{\vec{k}} \equiv \left[ \begin{array}{c}
c_{\vec{k} \uparrow} \\ c_{-\vec{k} \downarrow}^\dagger \end{array} \right]
\; ,
\end{equation}

\noindent the noninteracting part of the Gossamer Hamiltonian, the part with
the $U$ term disregarded, is

\begin{equation}
\text{$\cal{H}$} = \frac{1}{4}\frac{(2-\alpha_0)^2}{1-\alpha_0} \left[
\begin{array}{cc}
\epsilon_{\bf k}  & \Delta_{\bf k} \\
\Delta_{\bf k} &  - \epsilon_{\bf k}
\end{array} \right]
\end{equation}

\noindent where $\mu$ has been omitted because we are at half-filling.
The bare Green function, $G_{\bf k}(E)=1/(E - \text{$\cal{H}$})$, is then 
given by

\begin{equation}
G_{\bf k}(E)= \frac{1}{E^2 - \gamma^2 (\epsilon_{\bf k}^2
+ \Delta_{\bf k}^2 )} \left[ \begin{array}{cc}
E + \gamma \epsilon_{\bf k}  & - \gamma \Delta_{\bf k} \\
- \gamma \Delta_{\bf k} &  E - \gamma \epsilon_{\bf k}
\end{array} \right]
\end{equation}

\noindent with $\gamma \equiv (2-\alpha_0)^2 / 4 (1-\alpha_0)$.

In order to show the magnetic ordering properties of the Gossamer 
Hamiltonian
at half-filling, we will compute the magnetic susceptibility and tune it
through the transition. The bare susceptibility is given by

\begin{equation}
\chi_q^0 (\omega) = \frac{1}{(2\pi)^3} \int \int
{\rm Tr} [ G_k (E) G_{k+q}(E + \omega) ] \; dE dk \; \; .
\end{equation}

We calculate the effects of $U$ by the the ladder approximation for 
the the spin susceptibility:

\begin{equation}
\chi_q (\omega) = \frac{\chi_q^0 (\omega)}
{1 + U \chi_q^0 (\omega )} \; \; \; .
\label{rpa}
\end{equation}

\noindent The numerical evaluation of the spin susceptibility is shown in 
the
figure. We see that beyond a critical value for $U$ of order of $t$ and or
$\Delta_0$, the systems order antiferromagnetically, thus becoming an
antiferromagnetic insulator as signaled by the diverging susceptibility at
the critical value.

Being a somewhat crude approximation, the ladder technique will not provide
the right critical value of $U$ for which the transition to AFM order occurs,
nor will it provide the correct critical exponents. It will, however,
provide a faithful qualitative picture of the transition, of
the divergence of the spin susceptibility and the development of AFM
order. 

In the present note we demonstrated that, when extremely strongly projected,
the Gossamer superconductor has a continuous zero temperature phase transition
into an antiferromagnetic insulator. The Gossamer superconductor is
adiabatically continuable to a completely regular BCS superconductor. We can
thus conclude that superconducting matter at zero temperature undergoes
a transition to insulating antiferromagnetically ordered matter as the
short-ranged Coulomb repulsion is increased.

It was pointed out before\cite{bob} that under strong projection, the 
Gossamer
superconductor has a superfluid density that collapses with doping. This
collapsing superfluid density leads to a temperature order parameter
phase instability\cite{emery} consistent with the transition out of the
superconducting state in underdoped cuprates\cite{timusk}.
Even without the development of
antiferromagnetism, such a superconductor would be insulating since it would
dephase due to the small superfluid density\cite{dynes}.

We point out that, while arbitrarily small, the superfluid density is not zero 
for any doping including half-filling unless we project fully, something we
avoid as an uncontrolled approximation. Therefore the transition found here
can better be described as antiferromagnetic ordering developing underneath
the superconductivity\cite{hsu} thus opening an antiferromagnetic gap at the
d-wave node which is, in principle, observable in photoemission and tunelling
measurements.

An antiferromagnet with a small interpenetrating density of dephased superfluid
provides a possible explanation for the recent measurements of metallic 
transport below the N\'{e}el temperature in underdoped LSCO\cite{mottcond}. 
That
the charge mobility in these measurements is equal to that in the optimally 
doped material\cite{ando} suggests a common origin, possibly the dephased 
Gossamer superconductor.
Moreover, adding by hand an extra Hubbard term, an insulating static stripe
phase would be stabilized. The superconductor would undergo a transition 
similar to the one studied in the present letter into the stripe phase. 
Coupling of the coexisting dephased superfluid to the stripe phase would lead 
to anisotropic Copper-Oxygen plane charge transport\cite{ando1}. 

{\bf Acknowledgements} The authors thank Mac Beasley, Zaira Nazario,
Yidong Chong, Tanja Cuk, Brian Gardner, Alex Silbergleit and Joel Franklin for 
valuable help, comments and discussions.  This work was supported by DOE 
contract No. DE-AC03-76SF00515. B.A.B. also aknowledges the Stanford Graduate
Fellowship program support. D.I.S. was supported by NASA grant NAS 8-39225 to
Gravity Probe B.


\begin{thebibliography}{99}

\bibitem{phil} P. W. Anderson, Science{\bf 235}, 1196 (1987)

\bibitem{bob} R. B. Laughlin cond-mat/0209269 (2002).

\bibitem{fuji} T. Yoshida, {\it et. al.}, cond-mat/0206469 (Submitted
        to Phys. Rev. Lett.) (2002).

\bibitem{mottcond} Y. Ando and K. Segawa, Phys. Rev. Lett. {\bf 88},
        167005 (2002).

\bibitem{timusk} T. Timusk and B. Statt, Rep. Prog. Phys. {\bf 62}, 61
        (1999).

\bibitem{emery} V. J. Emery and S. A. Kivelson, Nature {\bf 374},
        434 (1995).

\bibitem{dynes} L. Merchant {\it et al.}, Phys. Rev. B {\bf 63},
        134508 (2001).

\bibitem{tranquada} J. M. Tranquada, {\it et. al.}, Nature {\bf 375},
        561 (1995).

\bibitem{rand} A. Paramekanti, {\it et. al.} Physical Review Letters {\bf 
87} 21002 (2001)

\bibitem{fz} F. C. Zhang cond-mat/0209272 (2002).

\bibitem{hsu} T. C. Hsu, Phys. Rev. B {\bf 41}, 11379 (1990).

\bibitem{ando} Y. Ando cond-mat/0206332 (to be published in Porceeding of the 
ICTP workshop) (2002).

\bibitem{ando1} Y. Ando, {\it et. al.}, Phys. Rev. Lett. {\bf 88}, 137005 
(2002).

\end{thebibliography}
\end{document}